\documentclass[a4paper]{spie}  %>>> use for A4 paper
\usepackage[dvips]{graphicx}

\title{ A dual output polarimeter devoted to the study of the Cosmic Microwave Background}

\author{Massimo Gervasi\supit{a}, Giuliano Boella\supit{a}, Francesco
Cavaliere\supit{b}, Giovanni Grossetti\supit{a}\supit{b}, \\
Andrea Passerini\supit{a}, Giorgio Sironi\supit{a}, Andrea
Tartari\supit{a}, and Mario Zannoni\supit{a}\supit{c}
\skiplinehalf
\supit{a}Universit\'a di Milano-Bicocca, Milan, Italy \\
\supit{b}Universit\'a di Milano, Milan, Italy \\
\supit{c}Istituto di Astrofisica Spaziale e Fisica Cosmica - CNR,
Milan, Italy}

\authorinfo{Further author information: (Send correspondence to Massimo Gervasi) \\
M. Gervasi: E-mail: massimo.gervasi@mib.infn.it, Telephone: +39 02
6448 2829; Address: Universit\'a di Milano-Bicocca - Physics
Department 'G. Occhialini', Piazza della Scienza 3, 20126, Milan,
Italy.}
\begin{document}
  \maketitle

%%%%%%%%%%%%%%%%%%%%%%%%%%%%%%%%%%%%%%%%%%%%%%%%%%%%%%%%%%%%%
\begin{abstract}
We have developed a correlation radiometer at 33 GHz devoted to
the search for the residual polarization of the Cosmic Microwave
Background (CMB). The two instrument's outputs are a linear
combination of two Stokes parameters (Q and U, or U and V). The
instrument is therefore directly sensitive to the polarized
component of the radiation (respectively linear and circular). The
radiometer has a beam-width of 7 or 14 deg, but it can be coupled
to a telescope increasing the resolution. The expected CMB
polarization is at most a part per milion. The polarimeter has
been designed to be sensitive to this faint signal, and it has
been optimized to improve its long term stability, observing from
the ground. In this contribution the performances of the
instrument are presented, together with the preliminary tests and
observations.
\end{abstract}

%>>>> Include a list of keywords after the abstract


\section{INTRODUCTION}
\label{sect:intro}  % \label{} allows reference to this section

The search for Cosmic Microwave Background (CMB) polarization is
one of the most promising to provide us with important information
about the primordial Universe. The current picture of the
Universe, as emerged from the CMB anisotropy measurements, can be
supported, enlarged, and clarified by the information coming from
the study of the CMB polarization. A residual linear polarization
of the CMB is expected from the presence of anisotropy. Its origin
comes from the anisotropic scattering of the CMB photons with the
cosmic plasma moving at the recombination epoch \cite{Rees68}. The
recent measurements of the CMB anisotropy power spectrum
\cite{DeBe00,Hanany00,DASI} improves the current knowledge of the
cosmic parameters, but it is unable to remove the degeneracy
between them. Therefore a measurements of the polarization power
spectrum can help to distinguish among scalar, vector and tensor
modes of density fluctuation.

The level of anisotropy measured sets the value of the expected
linear polarization to less than few $\mu K$ at degree angular
scale. The polarization signal is even fainter and decreasing at
larger scales \cite{HuWhite}. One of the most important answer we
can do through the polarization measurements is related to a
possible reionization occurred in the primordial Universe,
conditioning the post-recombination epoch. This information is
better obtained studying angular scales larger than 1 $deg$. If so
the polarization signal at these large scales is much larger and
comparable with the one expected at 1 $deg$ \cite{HuWhite}.
Finally deviations from a uniform and isotropic expansion of the
Universe produce a circular polarization on the CMB. A circular
polarization also appears in case of rotational modes associated
to primordial magnetic fields \cite{Kosowsky}. The present upper
limits, both linear and circular, are summarized in
Table~\ref{tab:uplim1} and Table~\ref{tab:uplim2}.

%% By placing this table in middle of paragraph,
%% it appears at bottom of page, but above the footnotes.
%% Use of [h] in following command forces table to appear "here".
\begin{table}[h]
\caption{Measured CMB polarization upper limits: linear
polarization.} \label{tab:uplim1}
\begin{center}
\begin{tabular}{|c|c|c|c|c|} %% this creates four columns
%% |l|l| to left justify each column entry
%% |c|c| to center each column entry
%% use of \rule[]{}{} below opens up each row
\hline
\rule[-1ex]{0pt}{3.5ex}
Wavelength ($cm$) & Angular Scale ($deg$) & Sky Region & Upper Limit ($\mu K$) & Reference \\
\hline \rule[-1ex]{0pt}{3.5ex} 3.4 & $3.8 \times 10 ^{-2}$  &
$\delta \simeq -50$ & 15 & Ref.\cite{Subra} \\
\rule[-1ex]{0pt}{3.5ex} 6.0 & $(5 - 44) \times 10 ^{-3}$ &
$\delta = +80$ & 93 - 23 & Ref.\cite{Partridge} \\
\rule[-1ex]{0pt}{3.5ex} 3.2 & 15 &
$\delta = +40$ & 1610 & Ref.\cite{Nanos} \\
\rule[-1ex]{0pt}{3.5ex} 1 - 0.75 & 2.9, 2.1, 1.6, 1.2, 0.9 &
$\delta = -63$ & 37, 54, 28, 41, 79 & Ref.\cite{Torbet} \\
\rule[-1ex]{0pt}{3.5ex} 0.91 & 7 - 15 & $ -37
\leq \delta \leq +63$ & 150 & Ref.\cite{LubinSmoot} \\
\rule[-1ex]{0pt}{3.5ex} 0.91 & 7 - 14 &
SCP & 267 - 226 & Ref.\cite{SironiASP} \\
\rule[-1ex]{0pt}{3.5ex} 1.2 - 0.8 & 1.2 &
NCP & 25 & Ref.\cite{Wollack} \\
\rule[-1ex]{0pt}{3.5ex} 0.33 & 0.85 &
$ \delta = +89$ & 14 & Ref.\cite{Headman} \\
\rule[-1ex]{0pt}{3.5ex} 0.05 - 0.3 & 0.5 - 40 & $ -10
\leq \delta \leq -45 $ & 300 & Ref.\cite{Caderni} \\
\rule[-1ex]{0pt}{3.5ex} 1.2 - 0.8 & 9 - 90 &
$ \delta = +43$ & 14 & Ref.\cite{Keating} \\
\hline
\end{tabular}
\end{center}
\end{table}

\begin{table}[h]
\caption{Measured CMB polarization upper limits: circular
polarization.} \label{tab:uplim2}
\begin{center}
\begin{tabular}{|c|c|c|c|c|}
\hline \rule[-1ex]{0pt}{3.5ex}
Wavelength ($cm$) & Angular Scale ($deg$) & Sky Region & Upper Limit ($\mu K$) & Reference \\
\hline \rule[-1ex]{0pt}{3.5ex} 6.0 & $(5 - 44) \times 10^{-3}$ &
$\delta = +80$ & 112 - 29 & Ref.\cite{Partridge} \\
\rule[-1ex]{0pt}{3.5ex} 0.91 & 15 & $ \delta = +37$ & $1.2 \times 10^{4}$ & Ref.\cite{Lubin} \\
\hline
\end{tabular}
\end{center}
\end{table}

The importance of the detection of the CMB polarization and the
very faint level of the signal make CMB polarimetry one of the
most exciting challenge for the next years in experimental
cosmology.

\section{THE RATIONALE OF THE INSTRUMENT}
\label{sect:rationale}

\begin{figure}
   \begin{center}
   \begin{tabular}{c}
   \includegraphics[height=5cm]{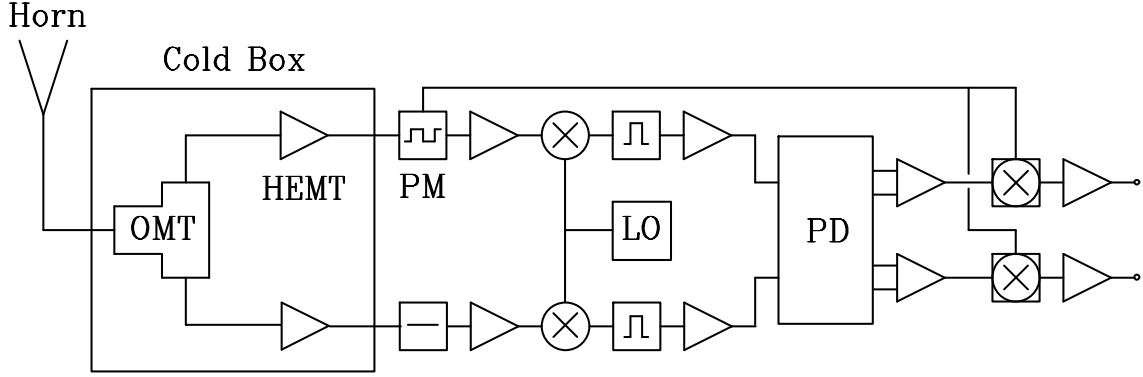}
   \end{tabular}
   \end{center}
   \caption[block diagram]
%>>>> use \label inside caption to get Fig. number with \ref{}
   { \label{fig:block}
A schematic of the Milano polarimeter. The main blocks are shown:
the cryogenic front-end, the modulation/demodulation loop, the
RF-to-IF down-conversion, the Phase Discriminator.}
   \end{figure}

We have developed a dual output polarimeter, operating at the
frequency of 33 $GHz$. The characteristics of the instrument are
summarized in Table~\ref{tab:char}, while the block diagram is
shown in Fig.~\ref{fig:block}. The instrument has been completely
described elsewhere \cite{SironiNA,SironiPASA}.

\begin{table}[h]
\caption{Main properties of the Milano polarimeter.}
\label{tab:char}
\begin{center}
\begin{tabular}{|c|c|}
\hline
\rule[-1ex]{0pt}{3.5ex} Frequency & 33 $GHz$ \\
\rule[-1ex]{0pt}{3.5ex} Bandwidth & 1.5 $GHz$ \\
\rule[-1ex]{0pt}{3.5ex} Angular Scale & 7 - 14 $deg$ \\
\rule[-1ex]{0pt}{3.5ex} Receiver & Heterodyne / Correlation \\
\rule[-1ex]{0pt}{3.5ex} Polarization modes & Linear / Circular \\
\rule[-1ex]{0pt}{3.5ex} Output & Dual / Stokes parameters \\
\hline
\end{tabular}
\end{center}
\end{table}

%-------------
%-------------

The radiation is fed by a corrugated conical horn into an
ortho-mode transduser (OMT). Here the two polarization components
are separated and conditioned in two distinct chains. The two
signals are then amplified, filtered, down-converted,
phase-modulated and finally correlated by a Phase Discriminator
(PD). The outputs of the PD are then detected synchronously with
the phase modulation.

The signals coming from the two chains are:

\begin{equation}
 {\bf A} = k_1 E_{01} e^{i \omega t} r(t)
\label{In1}
\end{equation}

\begin{equation}
 {\bf B} = k_2 E_{02} e^{(i \omega t + \gamma)}
\label{In2}
\end{equation}

Here $E_{0i}$ are the two polarization electric fields, $k_i$
accounts for the gain of the chains, $\gamma = \phi + \theta$ is
the phase difference at the $PD$ inputs, $r(t)$ describes the
square wave phase modulation on channel $A$. $\gamma$ is the
composition of the phase difference introduced by the instrument
($\theta$) with the intrinsic phase ($\phi$) of the radiation
components. The four outputs of the $PD$ are:

\begin{equation}
 U_1 = h_1 [|A|^2 + |B|^2 + 2AB \ r(t) \ \cos(\gamma)]
\label{Usc1}
\end{equation}

\begin{equation}
 U_2 = h_2 [|A|^2 + |B|^2 - 2AB \ r(t) \ \cos(\gamma)]
\label{Usc2}
\end{equation}

\begin{equation}
 U_3 = h_3 [|A|^2 + |B|^2 + 2AB \ r(t) \ \sin(\gamma)]
\label{Usc3}
\end{equation}

\begin{equation}
 U_4 = h_4 [|A|^2 + |B|^2 - 2AB \ r(t) \ \sin(\gamma)]
\label{Usc4}
\end{equation}

The parameters $h_i$ accounts for the different gains in the PD
ports. Generally we have $h_i \simeq h_j$ but the condition $h_i
\equiv h_j$ is not trivial to obtain. After a differential
amplification we have:

\begin{equation}
 S_1 = (h_1 - h_2) [|A|^2 + |B|^2] + (h_1 + h_2) 2AB \ r(t) \
 \cos(\gamma)
\label{Sig1}
\end{equation}

\begin{equation}
 S_2 = (h_3 - h_4) [|A|^2 + |B|^2] + (h_3 + h_4) 2AB \ r(t) \
 \sin(\gamma)
\label{Sig2}
\end{equation}

The common mode terms can be definitely rejected after the
synchronous detection\cite{Spiga}. Here in fact the signals are
demodulated: multiplied to a square wave $f(t) = \pm 1$
synchronous with $r(t)$ and integrated in a time interval much
longer than the modulation period. The resulting signals are:

\begin{center}
\begin{tabular}{l}
 $O_1 = < [(h_1 - h_2) [|A|^2 + |B|^2]] \times f(t) > +
 < (h_1 + h_2) 2AB \ \cos(\gamma) \ r(t) \times f(t) >$ \\
 \\
 $O_1 = \Gamma (h_1 + h_2) 2AB \ \cos(\gamma)$\\
 \\
 \\
 $O_2 = < [(h_3 - h_4) [|A|^2 + |B|^2]] \times f(t) > +
 < (h_3 + h_4) 2AB \ \sin(\gamma) \ r(t) \times f(t) >$ \\
 \\
 $O_2 = \Gamma (h_3 + h_4) 2AB \ \sin(\gamma)$
\end{tabular}
\end{center}

In this case the output signal is proportional to a linear
combination of two of the Stokes parameters, while $\Gamma \sim
<r(t) \times f(t)>$. Using the Iris polarizer we can study the
linear polarization:

\begin{equation}
 O_{1,l} = \Gamma (h_1 + h_2) 2AB \ \cos(\gamma) \propto Q \cos
 (\phi) - U \sin (\phi)
\label{Stokes1l}
\end{equation}

\begin{equation}
 O_{2,l} = \Gamma (h_3 + h_4) 2AB \ \sin(\gamma) \propto Q \sin
 (\phi) + U \cos (\phi)
\label{Stokes2l}
\end{equation}

Alternatively, without using the Iris polarizer, the system is
sensitive to the circular polarization:

\begin{equation}
 O_{1,c} = \Gamma (h_1 + h_2) 2AB \ \cos(\gamma) \propto Q \cos
 (\phi) - V \sin (\phi)
\label{Stokes1c}
\end{equation}

\begin{equation}
 O_{2,c} = \Gamma (h_3 + h_4) 2AB \ \sin(\gamma) \propto Q \sin
 (\phi) + V \cos (\phi)
\label{Stokes2c}
\end{equation}

%-------------
   \begin{figure}
   \begin{center}
   \begin{tabular}{c}
   \includegraphics[height=10cm]{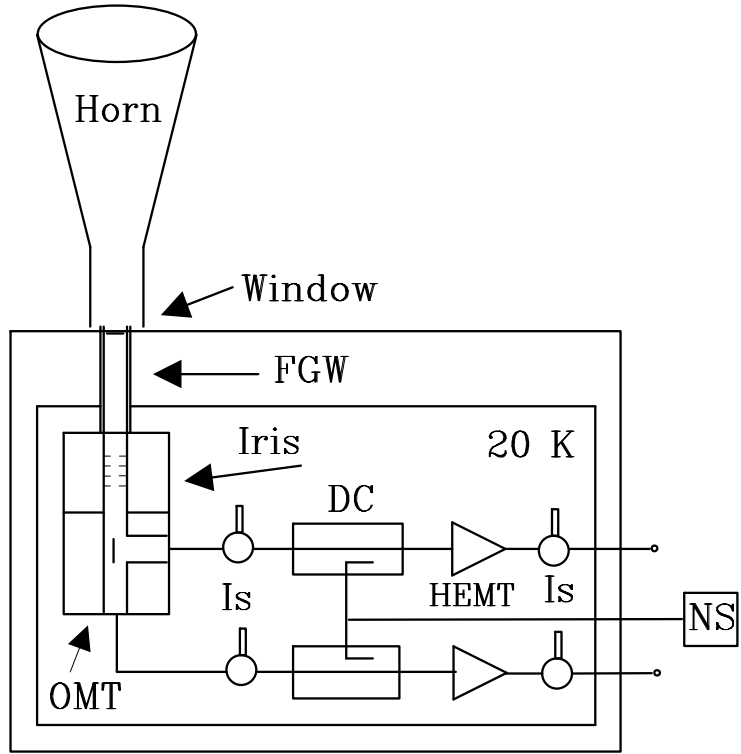}
   \end{tabular}
   \end{center}
   \caption[Front-end]
%>>>> use \label inside caption to get Fig. number with \ref{}
   { \label{fig:front-end}
Front-end section of the polarimeter: Horn, Polyethylene Window,
Fiber-glass Waveguide (FGW), Iris Polarizer, Ortho-mode transducer
(OMT), cryogenic isolators (Is), directional couplers (DC)
injecting the signal coming from the noise source (NS), and low
noise HEMT amplifiers. The cryogenic conponents are operated at $T
\sim 20 \ K$.}
   \end{figure}
%-------------

\subsection{Front-end and cryogenics}
\label{sect:front-end}

The front-end section is shown in Fig.~\ref{fig:front-end}. The
radiation from the horn comes through a circular fiberglass
waveguide (FWG) to the cryogenic components. The inner surface of
the fiberglass waveguide is silver + gold plated few $\mu m$ thick
in order to get a high electrical conductivity, while having a
good thermal insulation. The FGW is connected to the horn by a
polyethylene window. The Iris Polarizer is used to phase shift one
component of the electrical field against the other one. A phase
shift of $90^{\circ}$ is required to switch from linear to
circular components in the polarization. Now the radiation
components are separated by means of an Orthomode Transducer
(OMT). From this point we have two distinct chains of
amplification.

The cryogenic low noise amplifiers are High Electron Mobility
transistors (HEMT) provided by NRAO (noise temperature $T_N \simeq
10 \ K$, Gain $G \simeq 40 \ dB$, both measured at the temperature
of 15 $K$). We used also two pairs of isolators at the OMT outputs
and at the HEMT outputs. The calibration mark produced by a noise
source (NS) is introduced into the amplification chains at the
cryogenic section by two directional couplers (DC). The front-end
section, apart from the horn, is cooled down to 20 $K$ by a
mechanical cryocooler, reducing the insertion loss of the cold
components.

\subsection{RF and IF segments}
\label{sect:correla}

The RF section is extended at room temperature. The connection is
done by means of Stainless Steel WR-28 waveguides. These
waveguides are silver + gold plated in order to reduce the
insertion loss. The first component at room temperature are the
phase modulators (PM). These pin-diode devices change the phase of
the radiation switching between two states corresponding to $zero$
and $\pi$ phase difference. Only one of them is actually
switching, the other modulator is inserted with the purpose of
equalize the attenuation of the chains. Then we have a RF
amplifier and a band-pass filter selecting the RF band: $31.5 \
GHz \leq \nu \leq 34.5 \ GHz$.

\begin{figure}
   \begin{center}
   \begin{tabular}{c}
   \includegraphics[height=6cm]{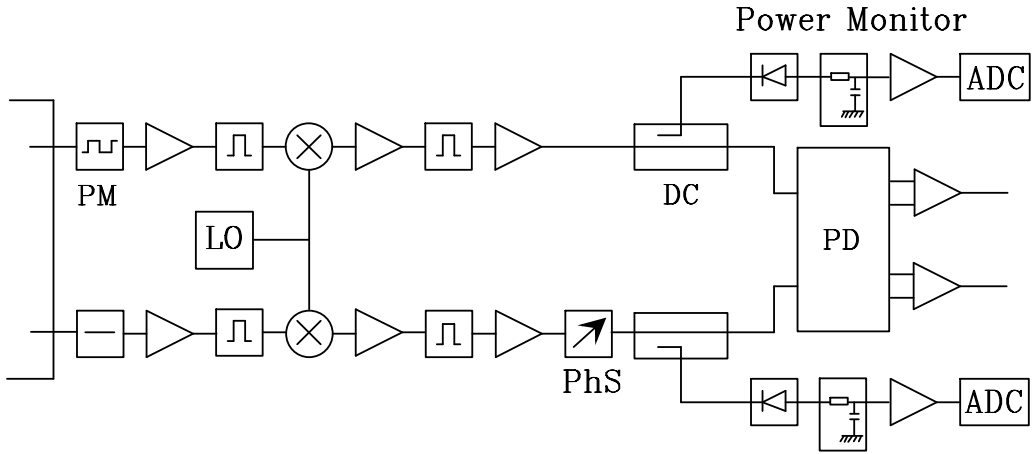}
   \end{tabular}
   \end{center}
   \caption[RF-IF]
%>>>> use \label inside caption to get Fig. number with \ref{}
   { \label{fig:RF-IF}
RF segment of the polarimeter: signal is phase modulated (PM),
amplified, RF filtered and down-converted using a common local
oscillator (LO). IF segment of the polarimeter: signal is
amplified and filtered again. A phase shifter (PhS) is added in
order to change the phase difference. A fraction of the signal is
extracted by two directional couplers (DC) and used as Power
Monitor. }
   \end{figure}

The down-conversion is driven by a common local oscillator (LO)
with a low phase noise at a frequency $\nu_{LO} = 30 \ GHz$ . The
IF frequency is therefore $\nu_{IF} = 3 \ GHz$. The IF section is
made by a couple of amplifiers and a band-pass filter selecting
the final band: $32.25 \ GHz \leq \nu \leq 33.75 \ GHz$. We insert
a variable phase shifter in one of the two chains in order to
change the phase difference helping the optimization of the
detector and for calibration purpose. Finally we derive a small
fraction of the amplified signal (1\%) and detect it as a power
monitor of the two chains. The RF and IF sections are shown in
Fig.~\ref{fig:RF-IF}.

%-------------
%-------------

\subsection{Correlation and synchronous detection}
\label{sect:modula}

\begin{figure}
   \begin{center}
   \begin{tabular}{c}
   \includegraphics[height=5cm]{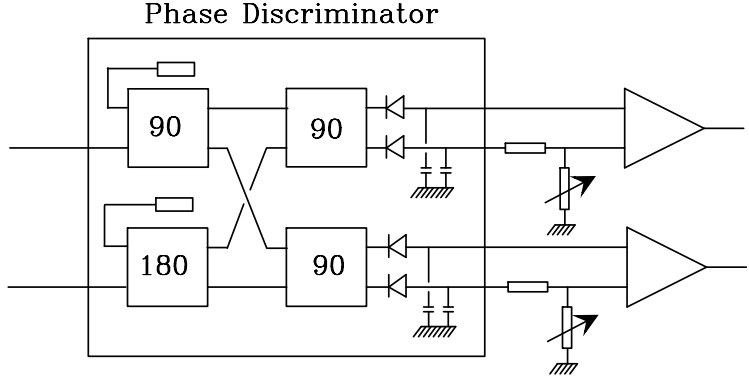}
   \end{tabular}
   \end{center}
   \caption[corr]
%>>>> use \label inside caption to get Fig. number with \ref{}
   { \label{fig:corr}
The phase discriminator: scheme of operation. The four hybrids are
shown. The IF signal is detected by four crystal diodes. The
amplification of the four signals is equalized by using a
resistive partition.}
   \end{figure}

\begin{figure}
   \begin{center}
   \begin{tabular}{c}
   \includegraphics[height=6cm]{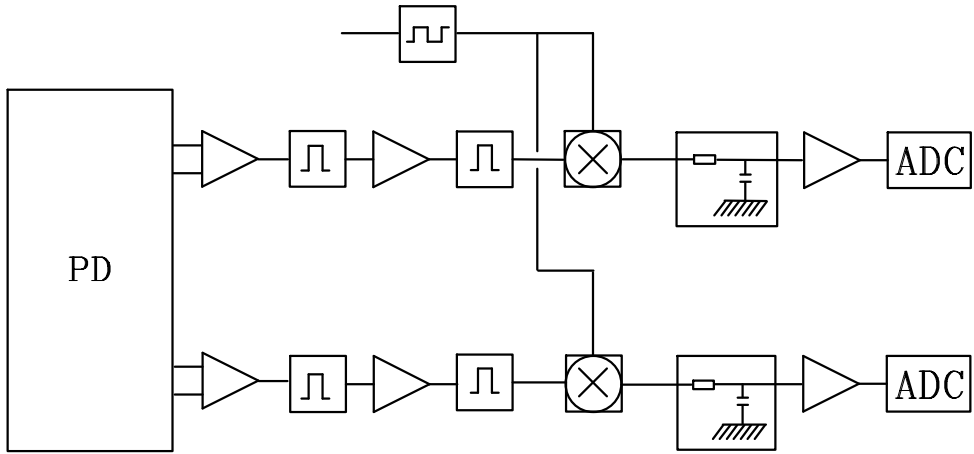}
   \end{tabular}
   \end{center}
   \caption[DC]
%>>>> use \label inside caption to get Fig. number with \ref{}
   { \label{fig:DC}
Post-detection segment of the polarimeter. The signal is filtered
at the modulation frequency and amplified before the synchronous
detection and the integration.}
   \end{figure}

The Phase discriminator (PD) is the core of the polarimeter. The
working scheme is shown in Fig.~\ref{fig:corr}. This device is
made by four hybrid couplers at 90 and 180 $deg$. They couple the
two inputs, as described in Fig.~\ref{fig:corr}. The four outputs
are now converted by four crystal detectors integrated into the
phase discriminator. At the PD output two differential amplifiers
extract the correlated signal, erasing the common mode. The gain
of the detection diodes are equalized through a resistive
partition, before entering the differential amplifier. The
described procedure is necessary to reduce the level of
systematics (see section~\ref{sect:systematics}). The two signals
are now filtered and amplified at the frequency of modulation.
Finally they are synchronously detected and then integrated. The
post-detection segment is shown in Fig.~\ref{fig:DC}.

%-------------
%-------------

%-------------
%-------------

\section{THE CALIBRATION PROCEDURE}
\label{sect:calib}

\subsection{The internal calibration mark}
\label{sect:NG}

\begin{figure}
   \begin{center}
   \begin{tabular}{c}
   \includegraphics[height=6cm]{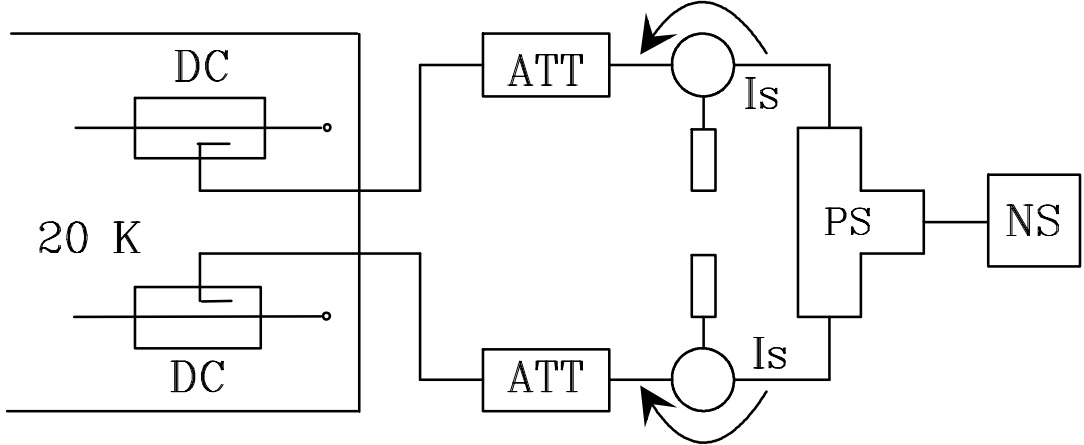}
   \end{tabular}
   \end{center}
   \caption[noise]
%>>>> use \label inside caption to get Fig. number with \ref{}
   { \label{fig:noise}
The internal calibration source. The signal from the noise source
(NS) is divided by a power splitter (PS). In each chain we have an
isolator (Is), a $30 \ dB$ attenuator (Att), and a directional
coupler (DC) used to inject the signal into the amplification
chains.}
   \end{figure}

We use a calibration mark, as shown in Fig.~\ref{fig:noise},
generated by a WR-28 Noise Source (NS) with an excess noise ratio
$ENR \simeq 12 \ dB$. The signal is divided by a waveguide power
splitter (PS) and then injected into the amplification chains
through a $20 \ dB$ directional coupler (DC). Before entering the
DC the signal is attenuated ($-30 \ dB$) and guided into a
stainless steel gold plated waveguides (WG). The two outputs of
the diplexer are isolated using two ferrite circulators (Is). In
this way we have a calibration correlated signal:

\begin{equation}
 T_{cal} = T_{NS} \ exp [-(\tau_{PS} + \tau_{Att} + \tau_{Is} +
 \tau_{WG} + \tau_{DC})]
\label{Tcal}
\end{equation}

\begin{center}
\begin{tabular}{l}
$ T_{cal} \sim 20 \ mK $
\end{tabular}
\end{center}

Unfortunately we also have a correlated signal when the NS is off.
This is a possible source simulating a polarized sky signal. This
spurious signal is at most the dummy load represented by the NS
termination ($T_0^{NS} \sim 300 \ K$), supposing to have a
negligible return loss:

\begin{equation}
 T_{spur} = T_{0}^{NS} \ exp [-( \tau_{PS} + \tau_{Att} + \tau_{Is} +
 \tau_{WG} + \tau_{DC})]
\label{Tspur}
\end{equation}

\begin{center}
\begin{tabular}{l}
$ T_{spur} \sim 1 \ mK $
\end{tabular}
\end{center}

This is a correlated off-set signal which can be easily taken into
account. What can be actually much more dangerous is the
fluctuation of this level. These fluctuations can be set only by a
thermal origin, in fact, when the NS is off, the source of
spurious signal is a completely passive device. It is therefore
extremely important to keep under control the temperature of the
source of correlated signal. For this purpose we thermalize the NS
and its case together with the diplexer at a temperature
fluctuating not more than $\pm 0.1 \ K$. In this way we have a
fluctuation on the correlated signal:
 $ \Delta T_{spur} < 1 \mu K $.

%-------------
%-------------

The sky signal can be evaluated from this mark by taking into
account the attenuation of the front-end of the radiometer (horn,
fiberglass waveguide, window, Iris polarizer, ortho-mode
transducer, and cryogenic isolators): $\tau_{FE} = \tau_{horn} +
\tau_{win} + \tau_{FGW} + \tau_{OMT} + \tau_{Iris} + \tau_{CIs}$.
The phase difference added by the NS path length
($\gamma^{\prime}$), respect to the phase of the signal entering
the horn ($\gamma$), has to be also taken into account. Finally we
get the following conversion factors ($\beta$) for the two
outputs:

\begin{equation}
 \beta_1 (K / ADU) = {T_{cal} \over O_{1,cal}} \
 {\cos(\gamma^{\prime}) \over \cos(\gamma)} \ e^{\tau_{FE}}
\label{beta1}
\end{equation}

\begin{equation}
 \beta_2 (K / ADU) = {T_{cal} \over O_{2,cal}} \
 {\sin(\gamma^{\prime}) \over \sin(\gamma)} \ e^{\tau_{FE}}
\label{beta2}
\end{equation}

The phase difference of the NS correlated signal can be easily
measured using the phase shifter placed in the amplification
chain. We use this internal calibrator injecting it periodically
in order to check for gain and phase long term fluctuations in the
system. The NS is not able to measure the phase $\gamma$. This
will be done using an external calibration source.

\subsection{The external calibration source}
\label{sect:griglia}

The external source is a grid polarizer placed in front of the
horn, at a distance far enough ($\sim 50 \lambda$) to stay in the
far-field region of the antenna. This grid has been realized with
thin gold coated copper wires $0.3 \ mm$ in diameter. The step of
the wires is $s = 0.6 \ mm$, therefore we get a transmission
coefficient of the perpendicular component of the radiation
similar to the reflectivity of the parallel component. Both these
coefficients are $\sim 99 \%$. The dimensions of the grid are $50
\ mm \times 50 \ mm$.

The correlated signal produced by the grid is large, but its
surface cover a small fraction ($\sigma \sim 19\%$) of the antenna
beam. The signal is the sum of the antenna temperature ($T_{sky} +
T_{Atm}$) transmitted by the grid and the noise temperature
($T_{N}^{Is}$) reflected back to the radiometer. In our case
$T_{N}^{Is} \sim 20 \ K$ is the temperature of the two cryogenic
isolators placed at the output of the OMT. If the two isolators
are at the same temperature, the correlated signal generated by
the calibrator is\cite{ZannoniPhD}:

\begin{equation}
  T_{grid} = \sigma [T_{sky} + T_{Atm} + T_N^{Is} ] \times \cos(\theta) \sin(\theta)
\label{Tgrid}
\end{equation}

Here $\theta$ is the angle between the grid's wires and the
principal planes of the OMT. If we rotate the grid we get the full
characterization of this calibration signal. We have already
tested this calibration system during the observations carried out
from Dome Concordia on the Antarctic plateau during the local
summer 1998-99\cite{ZannoniPhD,GervasiGIFCO}.

\subsection{The power monitors}
\label{sect:powermonitor}

In addition to the described calibrators we can also measure the
total power from the two amplification chains. These signals can
be derived just before the Phase Discriminator by using two $20 \
dB$ directional couplers (see Fig.~\ref{fig:RF-IF}). This
information is useful to take under control the gain fluctuation
of the RF and IF segments of the system.

Besides we can use these signals for calibration purpose and to
evaluate the system noise temperature. The total power outputs can
be calibrated using absorbers at two different temperature (liquid
nitrogen and room temperature). In this way we get the sensitivity
($K/ADU$) and the noise temperature of the total power channels.

\section{ANALYSIS OF SYSTEMATICS AND SENSITIVITY}
\label{sect:systematics}

An unavoidable requirement for an experiment devoted to the
measurement of the CMB polarization is the stability of the
performances. This is mandatory to guarantee the noise suppression
during long integrations. In particular for correlation receivers
the minimum detectable signal is related to the gain fluctuation
($\Delta G$) and to the offset ($T_{off}$) generated by the
instrument:

\begin{equation}
  \Delta T_{rms} = \sqrt{ {\alpha \ T^2_{sys} \over \Delta \nu \ \tau} +
  T^2_{off} \Bigl({\Delta G \over G} \Bigr)^2 + \Delta T^2_{off} }
\label{Trms}
\end{equation}

Here $\alpha$ is a numeric factor depending of the radiometer
type, $\tau$ is the integration time, and $\Delta \nu$ is the
bandwidth. The first term represent the white noise related to the
system temperature of the radiometer. The other terms are
degrading the system sensitivity, especially for long time
integrations. In fact while the first term is decreasing with time
the other contributions are expected to increase. Therefore gain
stabilization and offset reduction are necessary to improve the
sensitivity.

The offset signals arise mainly by non ideal performances of the
system. In a correlation receiver spurious correlated signals are
generated also in presence of uncorrelated radiation only. These
signals can be produced only in the common parts, where the
signals propagate in the same device. The places where this
happens are the antenna (horn, iris polarizer, ortho-mode
transducer) and the phase discriminator.

In a direct detection receiver into the second and third terms we
have to put $T_{sys}$ ($\Delta T_{sys}$) in place of $T_{off}$
($\Delta T_{off}$). The difference can be several order of
magnitudes: $T_{sys} / T_{off} \sim 10^4 - 10^5$. This simple
consideration suggests why a correlation receiver is intrinsically
more stable than a direct detection one.

\subsection{The antenna}
\label{sect:sysfrontend}

Several spurious terms cane be generated on the antenna. The
correlation introduced by the feed horn is responsible for an
offset term related to the anisotropy of the sky radiation
\cite{Carretti}. The signal is also proportional to the product of
the co-polar and cross-polar radiation pattern of the antenna. A
reasonable value of this product is $\sim 10^{-2}$, therefore this
term is completely negligible, when the main contributor of the
sky signal is the CMB.

A more serious effect comes from the cross correlation generated
by the iris polarizer and by the ortho-mode transducer. The offset
signal can be written as follows \cite{Carretti}:

\begin{equation}
  T_{off}^{Iris + OMT} = SP_{OMT} (T_{sky} + T_{Atm} + T_N^{Ant}) +
  SP_{Iris} \Bigl(T_{sky} + T_{Atm} + T_N^{Horn} - {T_0^{Iris} \over e^{-\tau^*_{horn}}} \Bigr)
\label{ToffOMT}
\end{equation}

\begin{equation}
  T_N^{Ant} = T_N^{Horn} + {T_N^{Iris} \over e^{-\tau^*_{horn}}} + {T_N^{OMT} \over
  e^{-\tau^*_{horn}} \ \vert S_{L} \vert^2} \label{Tant}
\end{equation}

\begin{equation}
  SP_{OMT} = 2 {\Re (S_{A1}S_{B1}^*) \over \vert S_{A1} \vert^2}
\label{SysOMT}
\end{equation}

\begin{equation}
  SP_{Iris} = {1 \over 2} \Bigl( 1 - {\vert S_{C} \vert^2 \over \vert S_{L}
  \vert^2} \Bigr)
\label{SysIris}
\end{equation}

There are two distinct terms. The first is related to the non
ideality of the OMT. In fact it is proportional to the cross
coupling term $S_{B1}$ (or $S_{A2}$). Here we suppose that the
diagonal terms of the scattering matrix are $S_{A1} \sim S_{B2}
\sim 1$, while the cross coupling terms are $S_{B1} \sim S_{A2} <<
1$. The second term is generated by the Iris polarizer. Here we
suppose $S_{L} \sim S_{C}$. This second term is not present when
the polarimeter is operated without Iris polarizer. The
contributions are also related to the noise temperature of the
antenna components ($T_N^{Horn}$, $T_N^{Iris}$, $T_N^{OMT}$), to
the external radiation field ($T_{sky}$, $T_{Atm}$), and to the
physical temperature of the Iris polarizer ($T_0^{Iris}$).

In the previous formulae we have defined: $ \tau^*_{horn} =
\tau_{horn} + \tau_{win} + \tau_{FGW} $; while $ T_N^{Horn} =
T_N^{feed} + T_N^{win} + T_N^{FGW} $. For our polarimeter we have
\cite{SironiNA}: $ T_N^{Horn} \sim 20 - 30 \ K $, while $
T_N^{Iris} + T_N^{OMT} \leq 1 \ K$, and $ T_{sky} + T_{Atm} \sim
10 - 15 \ K$. The cross correlation term of the OMT is $S_{B1}
\sim S_{A2} \sim -40 \ dB$. We also estimate to have the same term
from the Iris polarizer. Therefore the offset signal produced is:
$T_{off}^{Iris + OMT} \sim few \ mK$. These components can change
the physical temperature by about 1\%, and anyway we monitor this
variation and are able to correlate it with the observed offset
signal.

One more term ($T_{off}^{RL}$) arises from the return loss of the
antenna ($R_{Ant}$). The signal radiated from one of the ports of
the OMT can be partially reflected and enter into the other port
of the OMT. This signal is $T_{back} \sim T_N^{Is} +
I_{Is}T_{HEMT}$. Here $I_{Is} \sim -20 \ dB$ is the isolation of
the cryogenic circulators. At the Phase Discriminator this signal
is correlated with the radiation field propagating forward:
$T_{for} \sim T_{Ant} + T_{HEMT} + I_{Is} T_N^{Is}$. At the PD
output we obtain:

\begin{equation}
  T_{off}^{RL} \sim (T_{HEMT} + T_N^{Is}) \sqrt {I_{Is} \ R_{Ant} \ (SP_{OMT} + SP_{Iris})}
\label{ToffRL}
\end{equation}

This term can be potentially of the same order of magnitude of
$T_{off}^{Iris + OMT}$.

\subsection{Modulation and correlation}
\label{sect:syscorr}

Spurious signals can be also generated by the non ideal
performances of the modulation/demodulation system and of the
phase discriminator. We have seen as the combination of these
devices erases the offset signal due to the lack of cancellation
on the common mode terms\cite{Spiga}. The $AC$ detection technique
helps also to improve the stability of the instrument
performances, reducing the $1/f$ noise\cite{Spiga}. The frequency
of the phase modulation is $\nu_m = 781.25 \ Hz$. This frequency
has been chosen taking into account the major contributor of
electromagnetic pollution, i.e. the electrical power supply
network. Therefore we avoided to stay too close to the $50 \ Hz$
harmonics (and the 60 $Hz$ too)\cite{GervasiBo}.

The phase modulator is a pin-diode switch connecting input and
output through two different path lengths. The two paths are
selected reversing the polarization of the device. Both diodes and
micro-strip lines present an attenuation. If the attenuation
through the two ways is not the same an amplitude modulation is
introduced together with the phase one. Even if the amplitude
modulation is not so deep, the effect can be huge when an
unpolarized source is observed.

If an amplitude asymmetry ($\delta << 1$) is present, on the $A$
channel, as it is the modulated one, the output signals, after the
synchronous detection, are:

\begin{equation}
 O_1 = \Gamma [(h_1 - h_2) |A|^2 \delta + (h_1 + h_2) \ 2AB \ (1 +
 \delta/2) \ \cos(\gamma)]
\label{Mod1}
\end{equation}

\begin{equation}
  O_2 = \Gamma [(h_3 - h_4) |A|^2 \delta + (h_3 + h_4) \ 2AB \ (1 +
  \delta/2) \ \sin(\gamma)]
  \label{Mod2}
\end{equation}

Two terms are now added to the one taking the information of the
correlation. One of these terms is only a minor ($\delta << 1$)
correction of the correlated signal. The other effect is to
introduce a term depending on the common mode ($|A|^2$). This term
can be reduced if the gains at the Phase Discriminator outputs are
equalized, and the asymmetry in the amplitude modulation
($\delta$) minimized. The equalization of the outputs can be done
by means of the resistive partitions placed before the
differential amplifiers (see Fig.~\ref{fig:corr}). The modulation
asymmetry $\delta$ can be minimized by fine tuning the pine-diodes
bias amplitude.

This spurious term enters into the output signal as an offset, but
fluctuations in the system performances can mimic a correlation
signal. We have analyzed also displacements from the ideal
performances of the PD. Both amplitude asymmetries and phase
couplings at angles different from the $90^{\circ}$ and
$180^{\circ}$ induce only losses in the efficiency of the PD, but
no spurious signal is generated.

\section{OBSERVATION STRATEGY AND PROGRAMS}
\label{sect:observation}

The instrument has been tested at the top of the Physics
Department at the University of Milano - Bicocca for several
months. These measurements have been devoted to debug the
instrument from the systematics described previously testing the
solutions adopted. In order to get high sensitivity the instrument
performances have to be stable. This condition is required for
long time integration. The instrument is operated inside a
thermally insulating shelter. This solution has been adopted in
order to create a stable environment around the radiometer. In
this way we are able to thermalize the instrumentation and
suppress the gain fluctuation. The horn is looking at the sky
through a hole. We have selected an insulating tent operating down
to a temperature of $-70^{\circ}C$. In fact one of observing sites
selected for the polarimeter is the Antarctic plateau, during the
local winter. Therefore this solution is ideal also in places with
extreme climatic conditions. Besides the shelter is light and
easily transportable\cite{SironiPASA}.

The polarimeter will start celestial observations from Testa
Grigia at the Plateau Ros\`a \cite{MITO1} on the Italian Alps, at
3480~m a.s.l. The instrument will be installed looking directly at
the sky. Therefore we will observe with an angular resolution of
$7^{\circ}$ and $14^{\circ}$. Observations will be carried out in
transit mode, avoiding any movement of the instrumentation and of
the beam direction. This can help for reducing the fluctuation of
the signal coming from the local sources (ground and atmosphere).
Then the polarimeter will be placed at the focal plane of the MITO
telescope. In this way we get an angular resolution of $10-20 \
arcmin$ {\cite{MITO1,MITO2}.

In a second time the polarimeter will be operated from the
Antarctic Plateau in order to take advantage of the long Antarctic
night. There observations in quasi static weather conditions are
possible, avoiding solar radiation disturbances. Natural targets
are the South Celestial Pole (SCP) regions. We can cover circles
as large as few beam size around the pole. Besides, when observing
the SCP, in case of linear polarization a typical signature is
expected:

\begin{equation}
  O_1 \propto \sin(4 \pi {t \over T} + \phi_{\circ})
\label{SCP1}
\end{equation}

\begin{equation}
  O_2 \propto \cos(4 \pi {t \over T} + \phi_{\circ})
\label{SCP2}
\end{equation}

Here $t$ is the sidereal time, and $T$ a sidereal day, while
$\phi_{\circ}$ is an arbitrary phase angle.

%%%%%%%%%%%%%%%%%%%%%%%%%%%%%%%%%%%%%%%%%%%%%%%%%%%%%%%%%%%%%
\acknowledgments     %>>>> equivalent to \section*{ACKNOWLEDGMENTS}

This research was carried out within the Concordia Project
(supported by IFRTP and PNRA), and has been supported by PNRA and
CSNA, CNR, MURST and the Universities of Milano and Milano
Bicocca.

%%%%%%%%%%%%%%%%%%%%%%%%%%%%%%%%%%%%%%%%%%%%%%%%%%%%%%%%%%%%%
%%%%% References %%%%%

\bibliography{art_pol}   %>>>> bibliography data in art_pol.bib
\bibliographystyle{spiebib}   %>>>> makes bibtex use spiebib.bst

\end{document}